\documentstyle[11pt,appb]{article}
\def \seff {s^2_{eff}}
\def \ms   {\overline{\mbox{MS}}}
\newcommand{\be}{\begin{equation}}
\newcommand{\ee}{\end{equation}}
\newcommand{\bea}{\begin{eqnarray}}
\newcommand{\eea}{\end{eqnarray}}
\newcommand{\smallz}{{\scriptscriptstyle Z}} 
\newcommand{\smallw}{{\scriptscriptstyle W}} %
\newcommand{\smallh}{{\scriptscriptstyle H}} %
\newcommand{\mz}{M_\smallz}
\newcommand{\mw}{M_\smallw}
\newcommand{\mh}{M_\smallh}
\def \mt   {m_t}
\def \gev  {\mbox{ GeV}}
\newcommand{\equ}[1]{Eq.~(\ref{#1})}
\begin{document}
\begin{titlepage}
\begin{flushright}
        \small
        DFPD-98/TH/33\\
        hep-ph/9807293\\
        July 1998
\end{flushright}

\begin{center}
\vspace{1cm}
\renewcommand{\thefootnote}{\fnsymbol{footnote}}
\setcounter{footnote}{0}

{\large\bf Accurate prediction of electroweak observables and impact on the
       Higgs mass bound}\footnote{Talk presented at the Zeuthen Workshop
on Elementary Particle Theory "Loops and Legs in Gauge Theories"
Rheinsberg, Germany, April 19-24, 1998}

\vspace{0.5cm}
{\bf Giuseppe Degrassi}
\vspace{.8cm}

{\it Dipartimento di Fisica, Universit\`a di Padova, 
         Sezione INFN di Padova \\
         Via F.~Marzolo 8, I-35131 Padua, Italy}
\vspace{1cm}

{\large\bf Abstract}

\vspace{.5cm}
\end{center}
I discuss the importance of the $O(g^4 m_t^2/\mw^2)$ corrections to
the effective electroweak angle and $\mw$ in the indirect determination
of the Higgs mass. I emphasize the r\^ole of a very precise $\mw$
measurement on the $\mh$ estimate. 
\noindent
\end{titlepage}
\section{Introduction}
One of greatest achievement of the program of accurate verification of the
Standard Model of electroweak interaction (SM) carried out at LEP and SLC
during  the last decade, has been the prediction of the top mass. After 
the experimental discovery of the top quark by the CDF collaboration 
(with a mass at the right place indicated by the electroweak fits) the
challenge of precision physics has moved towards the only remaining unknown
particle of the SM, namely the Higgs. However, in this case the game is
much harder. The reason is clearly connected to the different behavior
of the virtual effects of the two particles in the relevant electroweak
corrections: power-like for the top, much milder and just logarithmic for the 
Higgs. To appreciate how much this logarithmic behavior makes hard the game 
for the theorists (and the experimentalists  also) I consider the effective 
electroweak mixing angle, $\sin^2 \!\theta^{lept}_{eff} \equiv \seff$,
that is the most important quantity in the determination  of $\mh$, and 
write it as 
\be
\seff \sim (c_1 + \delta c_1) + (c_2 + \delta c_2) \log \, y; \quad
 y \equiv (\mh/100\, \gev) .
\label{e1}
\ee
In \equ{e1} I identify  the l.h.s.~with the experimental result
that, I assume, carries no error. In the r.h.s.~$\delta c_i$ represent the 
theoretical uncertainty in the corresponding coefficients connected to the fact
that we have computed $c_i$ in perturbation theory through  certain order
in the perturbative series and therefore we do not know their exact values 
because of higher order contributions.
From \equ{e1}\ one obtains 
\be
y = y_0\, exp  \left[ -  \frac{\Delta_{th}}{c_2} \right] \quad
\Delta_{th} =  \delta c_1 + \delta c_2 \log\, y
\label{e2}
\ee
where $y_0$ is the value corresponding to $\delta c_1= \delta c_2 =0$.
To see the effect of $\Delta_{th}$ in extracting 
$\mh$ I take  
\be
\quad c_2 \sim \frac{\alpha}{2  \pi(c^2 - s^2)}
\left(\frac56 - \frac34 c^2 \right) \sim 5.5 \times 10^{-4};
\quad \Delta_{th} \sim  \pm 1.4 \times 10^{-4}
\label{e3}
\ee
where $s^2  \sim 0.23, \:
c^2 = 1 - s^2$. In \equ{e3} I estimate
$c_2$ through the Higgs leading behavior of the 
correction $\Delta \hat{r}$ relevant for $\seff$ \cite{si89,DFS}
while for $\Delta_{th}$, 
I take the value estimated in the 1995 CERN report
on `Precision calculation for the Z resonance' \cite{CERN}.  
The latter has been obtained
comparing the output of five different codes that implement different 
renormalization schemes and have built in several options for resumming
known effects. At the time of the report, the knowledge of the electroweak
part of the radiative corrections included, besides the complete one-loop
order, the leading logarithms of 
$O(\alpha^n \log^n \mz/m_f)$ (here $m_f$ is a generic fermion mass)
\cite{AM,Si84}, the $O(\alpha^2 \log \mz/m_f)$\cite{Si84} term while
for the two-loop top contribution only the leading  $O(g^4 m_t^4/\mw^4)$
correction was known \cite{bar}. Therefore the comparison of the 
various codes was mainly measuring the scheme-dependence error induced by the 
ignorance of the next term in the two-loop top contribution, namely the
$O(g^4 m_t^2/\mw^2)$ corrections. Inserting the values of \equ{e3} into
\equ{e2} yields $y \sim 1.29 \,y_0$. We see that a theoretical uncertainty
coming from two-loop unknown contributions (that are supposed to be not
even the dominant part) makes an error in the indirect determination of
the Higgs mass  of 29 \%!
\section{Recent advance in higher order calculations}
The above example clearly tells us that to extract accurate indirect 
information on the Higgs one needs not only very precise experiments but
also a very good control of the theory side. This brings in the issue of
what  error we can associate to our theoretical predictions.
They are affected by  uncertainties  coming from two 
different sources: one that is called parametric and it is connected to the
error in the experimental inputs used in our predictions. The second one 
is called intrinsic and it is related to the fact that our knowledge of the
perturbative series is always limited, usually to the first few terms. 
Concerning 
parametric uncertainties, $\alpha(0), \, G_\mu$ and $\mz$ are very well
measured, $\mt$ and $\alpha_s$ are not so precisely known while for
$\mh$ there is not at all  direct evidence. The scale of the weak interactions
is given by the mass of the intermediate vector bosons, so what actually 
matters in our predictions is not $\alpha(0)$ but $\alpha(\mz)$. The latter
contains the hadronic contribution to the photon vacuum polarization,
$(\Delta \alpha)_h$,
that cannot be evaluated in perturbation theory. Fortunately, one can use a 
dispersion relation to relate it  to the experimental
data on the cross section for $e^+ e^-$ annihilation into hadrons. 
In the recent years
there has been a lot of activity on this subject. Several new analyses
 appeared that differ  in the treatment of the experimental data 
\cite{Yeg,Swa}
and in the amount of theoretical input used to evaluate them
\cite{MZ,Al}. The situation
is not yet settle down (and probably will not be till new experimental data
on the $e^+ e^-$ cross section in the low and intermediate energy region
are available), so a conservative approach is  still to use the
value given by the most phenomenological analyses \cite{Yeg},
$
\alpha(\mz)^{-1} = 128.90 \pm 0.09 .
$

The status of the intrinsic uncertainties has actually improved since the CERN
report. A sizeable amount of work on radiative corrections has been completed
in the recent past. In this talk, I will discuss only the information that
is now available on the $O(g^4 m_t^2/\mw^2)$ corrections. 

The fact that the top is heavier
than  the other known particles  suggests to organize  its two-loop
contribution to the various radiative parameters as a series in $\mt$. 
The first two terms of this series are enhanced by factors $(\mt^2/\mw^2)^n\,
(n=1,2)$ while the remaining ones are at most logarithmic in nature. 
The leading
contribution that scales as $\mt^4$ is  completely available for arbitrary
value of the Higgs mass since few years \cite{bar}.  The 
next term, i.e.~the $O(g^4 m_t^2/\mw^2)$ correction, has been recently 
incorporated in 
the theoretical calculation of $\mw$\cite{DGV}, $\seff$\cite{DGS,paolo} and
the partial widths of the Z into fermions but the $b$ quark \cite{DG}. Indeed 
in the case of the $b$  there are specific vertex corrections of the same
order not yet computed. To gauge the residual scheme dependence, $O(g^4)$, this
incorporation has been performed in three electroweak resummation 
approaches and two different ways of implementing the relevant QCD
corrections \cite{DGS}. One of the approaches ($\ms$) employs
$\hat{\alpha}(\mz)$ and $\sin^2 \!\hat{\theta}_{\smallw}(\mz) \equiv
\hat{s}^2$, the $\ms$ QED and electroweak mixing parameters
evaluated at the scale $\mu = \mz$, while the other two (OSI and OSII)
make use of the on-shell parameters $\alpha$ and 
$\sin^2 \!\theta_{\smallw} \equiv 1-\mw^2/\mz^2$. As expected, the
dependence on the electroweak scale $\mu$ cancels through 
$O(g^4 \mt^2/\mw^2)$. However, because complete $O(g^4)$ corrections
have not been evaluated, the $\ms$ and OSI formulations contain a
residual $O(g^4)$ scale dependence. On the other hand OSII is, by 
construction, strictly $\mu$-independent. In table 1 the predictions for
$\seff$ and $\mw$ in this three different frameworks  are shown. The QCD  
corrections are implemented on the base of a top pole mass parameterization 
(for results with QCD corrections implemented in terms of running
$\ms$ top mass see Ref.\cite{DGS}). For each entry
of the Higgs mass the first row corresponds to the value obtained
including only the $O(g^4 m_t^4/\mw^4)$ contribution while the second one
contains also the $O(g^4 m_t^2/\mw^2)$ part.
\renewcommand{\arraystretch}{1.1}
\begin{table}[t]
\label{tab1}
\[
\begin{array}{|c| r r r | r r r|}\hline
 & \multicolumn{3}{|c|}{\sin^2 \theta_{eff}^{lept}} 
& \multicolumn{3}{|c|}{\mw ({\rm GeV})}\\\hline 
\mh  &  {\rm OSI} & {\rm OSII} 
& \rm{\overline{MS}} 
&  {\rm OSI} & {\rm OSII} & \rm{\overline{MS}} 
  \\  \hline\hline
65   & .23131 & .23111 & .23122 & 80.411 & 80.422 & 80.420 \\ 
     &     32 &     34 &     30 &     05 &     04 &     06 \\ \hline
100  & .23153 & .23135 & .23144 & 80.388 & 80.397 & 80.396 \\ 
     &     53 &     55 &     52 &     82 &     81 &     83 \\ \hline
300  & .23212 & .23203 & .23203 & 80.312 & 80.316 & 80.319 \\ 
     &     10 &     14 &     10 &     08 &     06 &     08 \\ \hline
600  & .23251 & .23249 & .23243 & 80.256 & 80.257 & 80.263 \\
     &     49 &     52 &     49 &     54 &     52 &     54 \\ \hline
1000 & .23280 & .23282 & .23272 & 80.215 & 80.213 & 80.221 \\
     &     77 &     79 &     77 &     14 &     13 &     14 \\
\hline
\end{array}            
\]
\caption{Predicted values of $\mw$ and $\seff$ in different frameworks
for $\mt = 175$ GeV with QCD corrections based on pole top-mass 
parameterization. The first row of each $\mh$ entry is obtained 
including only the $O(g^4 \mt^4/\mw^2)$. The $O(g^4 \mt^2/\mw^2)$ 
result is presented in the  second row (only the last two different
digits are shown).  }
\end{table}
I will not discuss in detail the effect of the $O(g^4 \mt^2/\mw^2)$ corrections
in the electroweak fits (see Bob Clare's talk \cite{Clare}) but I would like
to point out few things that can be easily read from table 1.
i) The incorporation of the $O(g^4 \mt^2/\mw^2)$ corrections  reduces
the scheme dependence to the level of $4 \times 10^{-5}$ in $\seff$
and $2$ MeV in $\mw$. ii) The $O(g^4 \mt^2/\mw^2)$ values for $\seff$
($\mw$) are generally higher (lower) than the corresponding 
$O(g^4 \mt^4/\mw^4)$ results. In the indirect determination of $\mh$ this fact
favors a lighter value of the mass. iii) In general the $O(g^4 \mt^2/\mw^2)$
OSI and $\ms$ results are very close. The OSI resummation is 
actually the natural generalization  to
$O(g^4 \mt^2/\mw^2)$ of the one proposed by Consoli-Hollik-Jegerlehner 
\cite{CHJ}
for the reducible $O(g^4 \mt^4/\mw^4)$ term and it  is the one presently
implemented in ZFITTER \cite{ZFit}. On the other side our $\ms$ 
approach \cite{DFS} is quite similar to the one implemented  in TOPAZ0 
\cite{TOP}.
This explain why in the new version of the 
famous LEPEWWG $\Delta \chi^2$ vs.~$\mh$ blue-band plot \cite{Clare} 
 the ZFITTER and TOPAZ0 lines are very close especially for large values of
$\mh$ and the blue band seems to have disappeared. With respect to this
a comment is in order. The  new $\Delta \chi^2$ curve obtained including
$O(g^4 \mt^2/\mw^2)$ corrections is not enclosed in the old blue-band 
representing the $O(g^4 \mt^2/\mw^2)$ scheme-dependence uncertainty 
\cite{EWWG}.
There is  nothing wrong with it. Indeed the comparison of results obtained in 
different schemes of calculation that contain all the available theoretical 
information at a given order of accuracy gives us  a guess of the
size of the reducible contribution, namely the part due to resummation or 
iteration of lower order effects. It does not tell us anything about the 
exact size of  higher order one-particle irreducible contributions. 
This way of estimating the intrinsic uncertainty should be taken as
giving  just the  order of magnitude of it and moreover  can be realistic 
only if the irreducible part is comparable or smaller
than the reducible one. But we have no way to know it before actually 
performing the calculation of  the irreducible part. 

\section{Importance of a precise $\mw$ measurement}
The precise electroweak measurements allow to constrain significantly the value
of the Higgs mass. A global fit to all data gives a strong indication for a
light Higgs with an upper limit at 95 \% C.L. $\mh < 215$ \gev\ \cite{Clare}.
However, the current estimates of $\mh$ depend crucially on the world average 
$\seff = 0.23149 \pm 0.00021$, and this follows from a combination of
experimental results that are not always in good harmony. 
The data presented at the recent
Winter conferences \cite{Clare} show a better agreement than the previous ones
\cite{EWWG}
but still the most precise LEP result ($\seff = 0.23213 \pm 0.00039$ from 
$A_{fb}^{0,b}$) and the SLAC data ($\seff = 0.23084 \pm 0.00035$) are quite 
far apart. To show how much the low value of SLAC is important for a
light $\mh$ determination I consider  $\seff$ and use the parameterization
\cite{DGPS}
\bea
\frac{\sin^2 \!\theta^{lept}_{eff}}{0.23151} -1 &=&
 b_1 \ln \left(\frac{\mh}{100\,\gev} \right) + 
 b_2 \left[ \frac{(\Delta \alpha)_h}{0.0280} -1 \right] \nonumber \\
&+ &  b_3 \left[ \left(\frac{\mt}{175 \,\gev} \right)^2 -1 \right]
  + b_4  \left[ \frac{\alpha_s(\mz)}{0.118} -1 \right]
\label{eq:s2}
\eea
that in  the range $75\gev \leq \mh \leq 350$ GeV, with the other parameters
within their $1-\sigma$ errors, approximates the 
detailed calculations of Ref.\cite{DGS} with average absolute deviations of 
$\approx 4\times 10^{-6}$ and maximum absolute deviations of
$(1.1-1.3)\times 10^{-5}$  depending on the scheme while outside the above 
range, the deviations increase reaching $(2.6-2.8)\times 10^{-5}$ for 
$\mh=600$ GeV (the values of the $b_i$ coefficients for the $\ms$ scheme
are presented in table 2). 
\renewcommand{\arraystretch}{1.1}
\begin{table}[t]
\label{tab2}
\begin{center}
\begin{tabular}{|r|r|r|c|} 
\hline 
         & $b_i~~~~~~$ & $ d_i~~~~~~$   & $| b_i/ d_i |$\\ \hline \hline 
$ i = 1$ & $ 2.26 \times 10^{-3}$ &$ -7.2 \times 10^{-4}$ &$  \sim 3.1$ \\ 
\hline
2 & $ 4.26 \times 10^{-2}$ &$ -6.4 \times 10^{-3}$ &$  \sim 6.6$ \\ \hline
3 & $ -1.20 \times 10^{-2}$ &$ 6.7 \times 10^{-3}$ &$  \sim 1.8$ \\ \hline
4 & $ 1.94 \times 10^{-3}$ &$ -1.1 \times 10^{-3}$ &$  \sim 1.8$ \\ \hline
5 &                       &$ -1.0 \times 10^{-4}$ &             \\ \hline
\end{tabular} 
\caption{Values in the $\ms$ scheme of  $b_i$ $(i=1-4)$ in \equ{eq:s2} 
        and $d_i$ $(i=1-5)$ in \equ{eq:mw} and their ratio. }
\end{center} 
\end{table} 
Employing in \equ{eq:s2} $\mt = 174.1 \pm 5.4$ \gev,
$\alpha_s(\mz) = 0.118 \pm 0.003$, $(\Delta \alpha)_h = 0.0280 \pm 0.0007$
and the LEP average for $\seff\: (\seff = 023186 \pm 0.00026)$ I obtain
a 95 \% C.L. upper bound $\mh < 610$ \gev. For the same values of
$\mt,\, \alpha_s(\mz)$ and $ (\Delta \alpha)_h$   the use of the SLAC value
for $\seff$ in \equ{eq:s2}\ 
gives instead a 95 \% C.L. upper bound  $\mh < 110$\gev. 
Clearly is the SLAC result that mainly pushes the electroweak fit towards a
light Higgs mass. Notice that a fit to LEP data alone (excluding the 
direct determination of $\mt$)  gives a light Higgs ($\mh = 56^{+101}_{-31}$)
but at the price of a low top ($\mt = 156^{+12}_{-10}$) \cite{Clare}. 
There is another observation to be made with respect to
$\seff$. This observable is very sensitive to  $ (\Delta \alpha)_h$. As I said,
the accuracy we know this quantity is presently under discussion. The most
conservative error \cite{Yeg} ($\delta (\Delta \alpha)_h = 7 \times 10^{-4}$) 
makes it the 
bottleneck in the improvement of the $\mh$ determination. The recent
more theoretically oriented analyses \cite{Al} give an error on 
$ (\Delta \alpha)_h$
ranging form $\delta (\Delta \alpha)_h = 1.6 \times 10^{-4}$
to $\delta (\Delta \alpha)_h = 4.5 \times 10^{-4}$. Using a smaller error for
$(\Delta \alpha)_h$ implies to weight more $\seff$ in the $\mh$ fit that
means we have to trust more the $\seff$ results.

This state of affairs strongly suggests the desirability of obtaining
constraints on $\mh$ derived from future precise measurements of $\mw$.
Similarly to \equ{eq:s2}\ I parameterize the result for  $\mw$ as \cite{DGPS}
\bea
\frac{\mw}{80.383} -1 &=&
 d_1 \ln \left(\frac{\mh}{100\,\gev} \right) + 
 d_2 \left[ \frac{(\Delta \alpha)_h}{0.0280} -1 \right] +
 d_5 \ln^2 \left(\frac{\mh}{100\,\gev} \right) \nonumber \\
&+ &  d_3 \left[ \left(\frac{\mt}{175 \,\gev} \right)^2 -1 \right]
  + d_4  \left[ \frac{\alpha_s(\mz)}{0.118} -1 \right]  
\label{eq:mw}
\eea
where  the $d_i$ coefficients are shown in table 2 and
notice that  to obtain an accuracy in the parameterization similar to
that of \equ{eq:s2}\ I need to introduce an extra term proportional to
$\ln^2 (\mh/100\,\gev)$. Comparing the coefficients of the \equ{eq:s2}
and \equ{eq:mw}\ we see that
at equal level of experimental accuracy (which is, in fact, the current 
situation)  $\seff$ is more
sensitive than $\mw$ by a factor $\approx 2.7$ in $\ln(\mh/100)$ (taking also
into account the $\ln^2(\mh/100)$ term of \equ{eq:mw}). On the other side,
$\mw$ has the welcome characteristic to be not so sensitive to  
$ (\Delta \alpha)_h$. 
\renewcommand{\arraystretch}{1.1}
\begin{table}
\label{tab3}
\begin{center}
\begin{tabular}{|c|c|c|} 
\hline 
\raisebox{-2ex}{$\ln (\mh/ (100\, \gev))$} & $\mw$ & $\seff $ \\
          & \raisebox{0.5ex}{determination} & \raisebox{0.5ex}{determination} 
\\ \hline \hline 
$\delta \mt = 3\, \gev, \, \delta \mw= 35$ MeV & & \\
\raisebox{0.5ex}{$\delta (\Delta \alpha)_h = 0.0007$} & \raisebox{+1.5ex}{
$ 0^{+0.663}_{-0.815}$} &  \raisebox{1.5ex}{$0 \pm 0.647$} \\ \hline
$\delta \mt = 1\, \gev, \, \delta \mw= 20$ MeV & & \\
\raisebox{0.5ex}{$\delta (\Delta \alpha)_h = 0.0007$} & \raisebox{+1.5ex}{
$ 0^{+0.404}_{-0.455}$} &  \raisebox{1.5ex}{$0 \pm 0.623$} \\ \hline
$\delta \mt = 1\, \gev, \, \delta \mw= 20$ MeV & & \\
\raisebox{0.5ex}{$\delta (\Delta \alpha)_h = 0.0002$} & \raisebox{+1.5ex}{
$ 0^{+0.352}_{-0.390}$} &  \raisebox{1.5ex}{$0 \pm 0.428$} \\ \hline
\end{tabular} 
\caption{Errors on $\ln  (\mh/ (100\, \gev))$ determined from $\mw$ 
        (\equ{eq:mw}) and $\seff$ (\equ{eq:s2})
        for $\mh = 100$ \gev, $\delta \seff = 0.00021$,  
        $\delta \alpha_s(\mz) = 0.003$ and different values of
        $\delta \mt$, $\delta \mw$ and $\delta (\Delta \alpha)_h$. }
\end{center}
\end{table} 
Let us now consider future scenarios where the 
experimental errors in the various quantities that enter in \equ{eq:s2}
and \equ{eq:mw} are somewhat reduced and compare the indirect determination
of $\mh$ from $\mw$ and $\seff$, separately. To make a simple comparison I
use central values that give the same Higgs mass, so I choose
$\seff = 0.23151, \, \mw = 80.383\, \gev, (\Delta \alpha)_h=0.0280, \, 
\mt = 175\, \gev, \, \alpha_s(\mz)=0.118$ that correspond to $\mh = 100$\gev. 
Table 3 presents 3 possible scenarios in all of which I assume no 
improvement in the $\seff$ and $\alpha_s (\mz)$ determination (i.e. $\delta 
\seff = 0.00021$ and $\delta \alpha_s (\mz) =0.003$) while the errors in
$\mt$, $\mw$ and in the last case also in $(\Delta \alpha)_h$ get reduced. 
One sees that a determination of $\mw$ at the level of 35 Mev together
with an improvement in $\mt$ to $\delta \mt = 3$\gev\, gives an
information on $\mh$ competitive with the one that is presently obtained
from $\seff$. Such a scenario is consistent with the expectation of Tevatron 
Run 2.
A further reduction in $\delta \mw$  and $\delta \mt$, that can be foreseen
at LHC, will make $\mw$  more effective than $\seff$ in determining $\mh$
even in a situation in which the error on $(\Delta \alpha)_h$ will be
significantly reduced.

\vspace{1.2cm}
I am grateful to  P.~Gambino, M.~Passera, A.~Sirlin
and A.~Vicini for the fruitful collaboration on the subject discussed here.
I would like also to thank T.~Riemann and the other organizers for the 
excellent organization and the pleasant atmosphere of the workshop.

\end{document}